\documentclass[10pt,twocolumn,letterpaper]{article}

\usepackage{cvpr}              

\usepackage{graphicx}
\usepackage{amsmath}
\usepackage{amssymb}
\usepackage{booktabs}

\usepackage{textcmds}

\makeatletter
\@namedef{ver@everyshi.sty}{}
\makeatother
\usepackage{pgfplots}

\usepackage{siunitx}
\usepackage{float}
\usepackage{tikz}

\usepackage{cuted}

\usepackage{adjustbox}
\usepackage{makecell, multirow}

\usepackage{algorithm}
\usepackage{algpseudocode}

\usepackage[pagebackref,breaklinks,colorlinks]{hyperref}

\usepackage[capitalize]{cleveref}

\crefname{section}{Sec.}{Secs.}
\Crefname{section}{Section}{Sections}
\Crefname{table}{Table}{Tables}
\crefname{table}{Tab.}{Tabs.}

\def\cvprPaperID{7551} 
\def\confName{CVPR}
\def\confYear{2022}

\begin{document}
\title{\ COSMOS: Cross-Modality Unsupervised Domain Adaptation for 3D Medical Image Segmentation based on Target-aware Domain Translation and Iterative Self-Training}

\author{Hyungseob Shin\thanks{indicates equal contribution.}, \ Hyeongyu Kim\footnotemark[1], \ Sewon Kim, \ Yohan Jun, \ Taejoon Eo, \ Dosik Hwang\\
School of Electrical and Electronic Engineering, Yonsei University\\
{\tt\small whatzupsup@yonsei.ac.kr}
}
\maketitle

  
%

\begin{abstract}
Recent advances in deep learning-based medical image segmentation studies achieve nearly human-level performance when in fully supervised condition. However, acquiring pixel-level expert annotations is extremely expensive and laborious in medical imaging fields. Unsupervised domain adaptation can alleviate this problem, which makes it possible to use annotated data in one imaging modality to train a network that can successfully perform segmentation on target imaging modality with no labels. In this work, we propose a self-training based unsupervised domain adaptation framework for 3D medical image segmentation named COSMOS and validate it with automatic segmentation of Vestibular Schwannoma (VS) and cochlea on high-resolution T2 Magnetic Resonance Images (MRI). Our target-aware contrast conversion network translates source domain annotated T1 MRI to pseudo T2 MRI to enable segmentation training on target domain, while preserving important anatomical features of interest in the converted images. Iterative self-training is followed to incorporate unlabeled data to training and incrementally improve the quality of pseudo-labels, thereby leading to improved performance of segmentation. COSMOS won the 1\textsuperscript{st} place in the Cross-Modality Domain Adaptation (crossMoDA) challenge held in conjunction with the 24th International Conference on Medical Image Computing and Computer Assisted Intervention (MICCAI 2021). It achieves mean Dice score and Average Symmetric Surface Distance of 0.871$\pm$0.063 and 0.437$\pm$0.270 for VS, and 0.842$\pm$0.020 and 0.152$\pm$0.030 for cochlea.

   
\end{abstract}

\section{Introduction}
\label{sec:intro}

\begin{figure}[htp]
\centering

\includegraphics[width=1\linewidth, keepaspectratio]{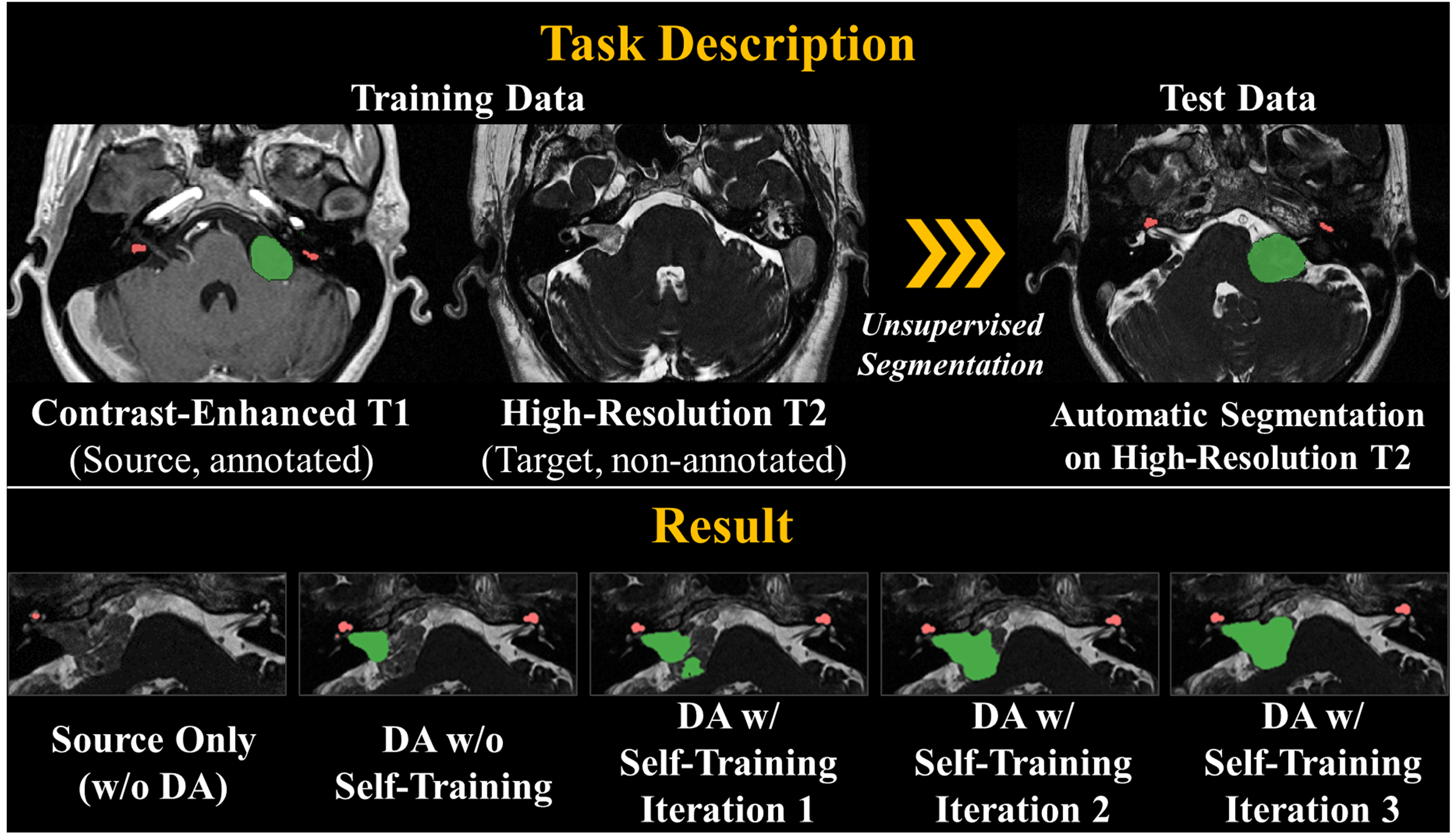}
\caption{An illustration that describes the task and result of this study. Top: Our COSMOS performs cross-modality unsupervised domain adaptation for 3D segmentation of VS (green) and cochlea (red) on unseen target domain given annotated source domain data and non-annotated target domain data. Source and target domain data were unpaired. Bottom: Segmentation results on target domain T2 MRI are provided. The segmentation performance improves as the proposed method is developed.}
\label{fig1}
\end{figure}

With the surprising development of deep learning, many studies are now showing remarkable performance in various applications. However, when a deep learning model faces data from an unseen domain, performance degradation occurs \cite{tzeng2017adversarial, ganin2015unsupervised}. Resolving this is important for the deep learning techniques to be applied in real world since collecting data from all domains and labeling them is very impractical and inefficient. Unsupervised Domain Adaptation (UDA) aims to make a model learned from one domain data work well in another domain, without the necessity of supervision in the target domain. 

Data dependency is more serious in medical image segmentation field since acquiring pixel-level expert annotation is extremely expensive and time-consuming \cite{chen2019synergistic, liu2019susan, zhang2018translating}. In this paper, we propose \textit{\textbf{C}ross-modality d\textbf{O}main adaptation for un\textbf{S}upervised 3D \textbf{M}edical image segmentation based on target-aware d\textbf{O}main translation and Iterative \textbf{S}elf-training (\textbf{COSMOS})} and validate it on 3D segmentation of Vestibular Schwannoma (VS) and cochlea on T2 Magnetic Resonance Imaging (MRI), as shown in \cref{fig1}.
\setlength \belowcaptionskip{-3ex}
\begin{figure*}[h!]
\centering
\includegraphics[width=2.0\columnwidth]{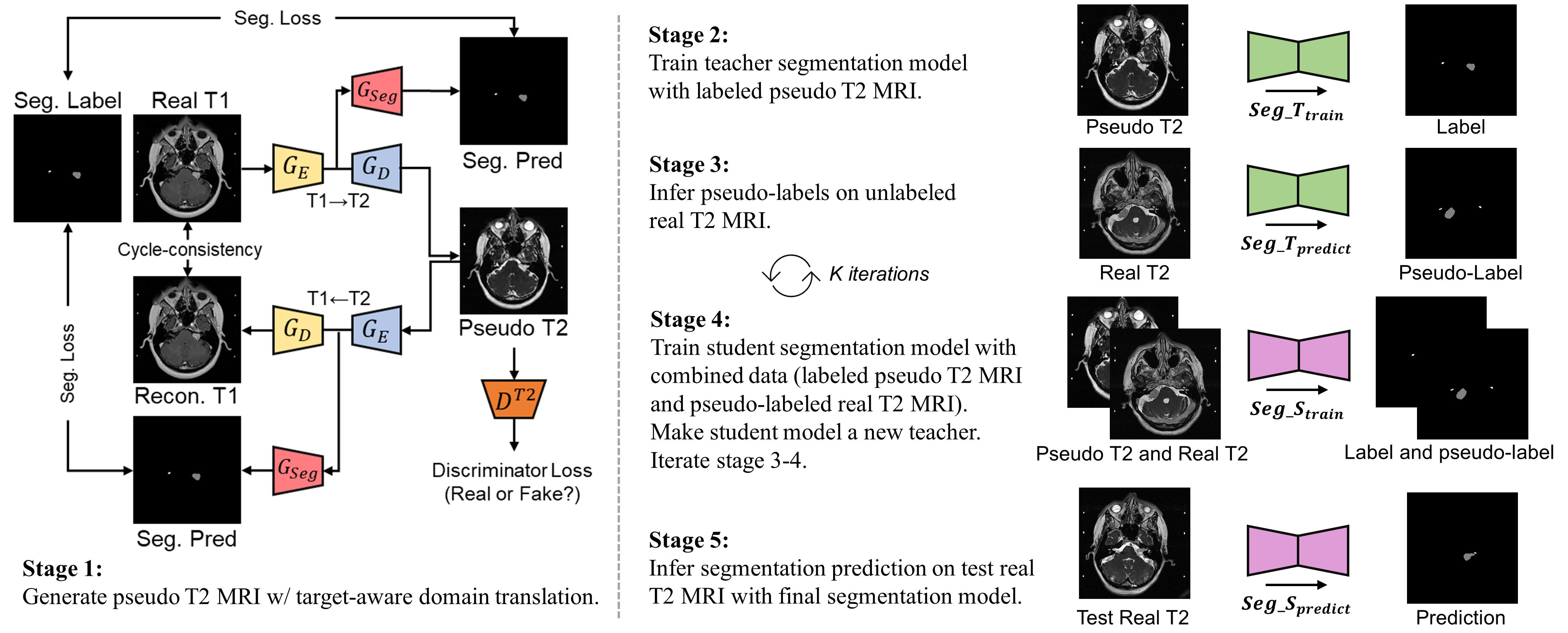}
\caption{Overview of our unsupervised cross-modality domain adaptation framework based on target-aware domain translation and iterative self-training. First, source-to-target image transformation is performed via target-aware domain translation that conducts domain translation and segmentation simultaneously, thereby preserving the detailed structures of VS and cochlea (stage 1). On the reverse loop, segmentation is not performed since we are assuming zero-label circumstance on target domain. The reverse loop is omitted for ease of illustration. Second, self-training is performed using combination of labeled pseudo T2 MRI and pseudo-labeled real T2 MRI (stage 2-4). Last, segmentation result is inferred on unseen target T2 MRI with the trained segmentation model (stage 5). Best viewed in color and on high-resolution display.}
\label{fig2}
\hfill
\end{figure*}

Fast and accurate diagnosis of VS is very important to prevent VS leading to hearing loss in clinical workflows, and the most typical MR protocol used or diagnosis has been a contrast-enhanced T1-weighted (ceT1) scan to observe VS and cochlea \cite{shapey2019artificial}. However, due to the potential side effect of Gadolinium injection and long scan time, segmenting VS and cochlea from high-resolution T2 (hrT2) scan is being studied as a substitute for ceT1 scanning to reduce the cost \cite{Reuben_scirbble}. We aim to perform 3D segmentation of VS and cochlea on hrT2 scans using unsupervised domain adaptation based on target-aware domain translation and iterative self-training. Our method utilizes unpaired sets of annotated ceT1 and non-annotated hrT2 scans to train a deep learning framework that performs automatic VS and cochlea segmentation on unseen hrT2 scans. Overall framework of the proposed method is illustrated in \cref{fig2}. \\

The main contribution of our work are as follows: 
\begin{itemize}
\itemsep0em 
    \item We present COSMOS, a novel cross-modality domain adaptation for unsupervised 3D segmentation based on target-aware domain translation and iterative self-training. 
    \item Target-aware domain translation: Target-aware domain translation performs both contrast conversion and segmentation simultaneously, thereby preserving anatomical features of interest (i.e., VS and cochlea) on generated pseudo images while successfully converting the global contrast.
    \item Iterative self-training scheme: We incrementally update the quality of pseudo-labels via self-training, which has not been validated in the medical image segmentation field, to the best of our knowledge. Improving pseudo-labels with self-training enables to train segmentation on target domain with confident supervision, resolving the performance degradation caused by domain gap.
    \item COSMOS is validated on the challenging task of unsupervised domain adaptation for 3D segmentation of VS and cochlea, outperforming other approaches with mean Dice score and Average Symmetric Surface Distance (ASSD) of 0.871$\pm$ 0.063 and 0.437$\pm$0.270 for VS, and 0.842$\pm$0.020 and 0.152$\pm$0.030 for cochlea.
\end{itemize}


\section{Related Work}
\label{sec:related}

\noindent \textbf{Unsupervised domain adaptation for medical image segmentation. }Unsupervised domain adaptation (UDA) for semantic segmentation is the most interested theme among UDA related researches \cite{hoffman2018cycada, kim2020learning, li2019bidirectional, tsai2018learning}. Especially, the concept of unsupervised domain adaptation on segmentation is very attractive in the field of medical imaging \cite{chen2019synergistic, Reuben_scirbble, huo2018synseg, jiang2020psigan, liu2019susan, zhang2018translating}, because it can address the difficulty of obtaining expert-level manual labels. 

Recent studies on UDA for medical image segmentation are mostly based on pixel-level adaptation in which domain translation and segmentation networks are trained end-to-end, thereby utilizing the source-to-target transformed images for training segmentation on target domain. \cite{liu2019susan,huo2018synseg,zhang2018translating } combined image-to-image translation and segmentation into a single network. \cite{chen2019synergistic} conducted adversarial training using the segmentation output of the synthetic/real target domain images to better preserve the geometry of the structure-of-interest. Moreover, \cite{jiang2020psigan} added both the synthetic/real target domain images and their corresponding segmentation probability maps to adversarial training to preserve not only the geometry of the structure-of-interest but also the appearance.

Our work differs from the previous works on the following aspects: First, COSMOS offers UDA for 3D medical image segmentation, whereas previous works on UDA for medical image segmentation are mostly based on 2D segmentation \cite{huo2018synseg, chen2019synergistic,jiang2020psigan }. \cite{zhang2018translating} tried 3D segmentation, but their method used manual labels on target imaging modality whereas COSMOS is assuming zero label situation on target imaging modality. Second, COSMOS can incorporate real target data into training and provide supervision to the target data with pseudo-labels created by self-training approach, which will be described in \cref{sec:subsec3_2}, whereas existing UDA methods bypass this through pixel-level or feature-level adaptation \cite{chen2019synergistic, huo2018synseg}. Last, COSMOS has been validated in the challenging task of segmenting small multi-class structures which are VS and cochlea, whereas previous works mainly focus on single-class segmentation or multi-class segmentation of large structures such as abdominal or cardiac structures.\cite{jiang2020psigan, liu2019susan}\\

\noindent \textbf{Self-configuring segmentation framework. }Many of the biomedical image segmentation approaches are limited to specific domains or conditions as it is extremely time-consuming and laborious to repeat the model configuration, training, and configuration tuning in a trial-and-error way to find the optimal solution for each data characteristic. 
On the other hand, self-configuring nnU-Net \cite{nnunet} automatically configures the preprocessing, network architecture, training, and post-processing steps for any new task by analyzing the data characteristics via pre-defined heuristic rules. \cite{nnunet} is specialized in medical image segmentation task, which first analyzes the input data fingerprint, configures model according to fixed and rule-based parameters, trains 2D and 3D segmentation models, and ensembles the results according to cross-validation performance with appropriate post-processing. \cite{nnunet} demonstrated the importance of the model configuration from preprocessing to post-processing and achieved state-of-the-art performance in 33 out of 55 medical image segmentation challenges with a simple U-Net architecture at the time of publication. In this study, we utilized nnU-Net as the backbone segmentation model for self-training.\\

\noindent \textbf{Self-training. }Self-training belongs to semi-supervised learning which has emerged to improve the resources and cost put into data labeling \cite{zoph2020rethinking, xie2020self}. It is widely used with the goal of improving the performance of classifiers by using lots of unlabeled data when there are few labeled data. 

In self-training, a teacher model is first trained using only the labeled data. Next, pseudo-labels with high confidence are inferred by passing the unlabeled data on the trained model. With the labeled data and pseudo-labeled data, a larger dataset can be used to train a student model that performs better than the teacher model trained only on labeled data. This process can be iterated by generating improved pseudo-labels with the updated model. Since self-training framework can incorporate unlabeled images into training, it is reported that it increases the model\textquotesingle s general robustness against out-of-distribution data \cite{xie2020self}. To this end, numerous attempts were made to apply self-training into semantic segmentation  \cite{zou2020pseudoseg , pastore2021closer ,zou2018unsupervised, zhu2020improving }, as pixel-level annotations are expensive.

It is a recent trend to combine self-training and domain adaptation together, to remedy the unsupervised image segmentation performance in zero-label target domain \cite{zou2018unsupervised ,liu2021cycle}. However, to the best of our knowledge, COSMOS is the first to apply self-training on unsupervised domain adaptation with medical segmentation, which can alleviate the domain discrepancy and lack of data in medical imaging at the same time.

\section{Methods}
\label{sec:formatting}

\subsection{Target-aware domain translation network}
\label{subsec:3_1}
Provided with unpaired datasets of two different MRI contrasts which are annotated source domain T1 scans \\ $ x^s = \{ x^s_i \}_{i=1}^{N_s} $, its annotations $ y^s=\{ y^s_i \}_{i=1}^{N_s} $, and non-annotated target domain T2 scans $ x^t=\{ x^t_i \}_{i=1}^{N_t} $, where $N_s = N_t= 105$, we first convert the source T1 scans to synthetic (i.e., pseudo) target T2 scans $ \tilde{x}^t=\{ \tilde{x}^t_i \}_{i=1}^{N_s} $ using CycleGAN-based network \cite{zhu2017unpaired}. It is to use the synthetic T2 images paired with the annotations on source T1 images, termed \qq{labeled pseudo T2 scans}  hereafter, in training a segmentation model on target T2 domain. 

\begin{equation}
CycleGAN^{T1\rightarrow T2}: x^s\rightarrow \tilde{x}^t \nonumber
\end{equation}

However, it is difficult to preserve the detailed geometrical structures-of-interest in the synthetic images with cycle-consistency alone \cite{cohen2018distribution}. We resolved this problem simply by attaching additional decoder for segmentation, termed segmentor, in parallel to the decoder for translation to perform domain translation and segmentation at the same time, thereby enforcing the shared-encoder to focus more on the segmentation area while compressing the features (see stage 1 of \cref{fig2}). 

\begin{figure*}[h!]
\centering
\includegraphics[width=1\linewidth, keepaspectratio]{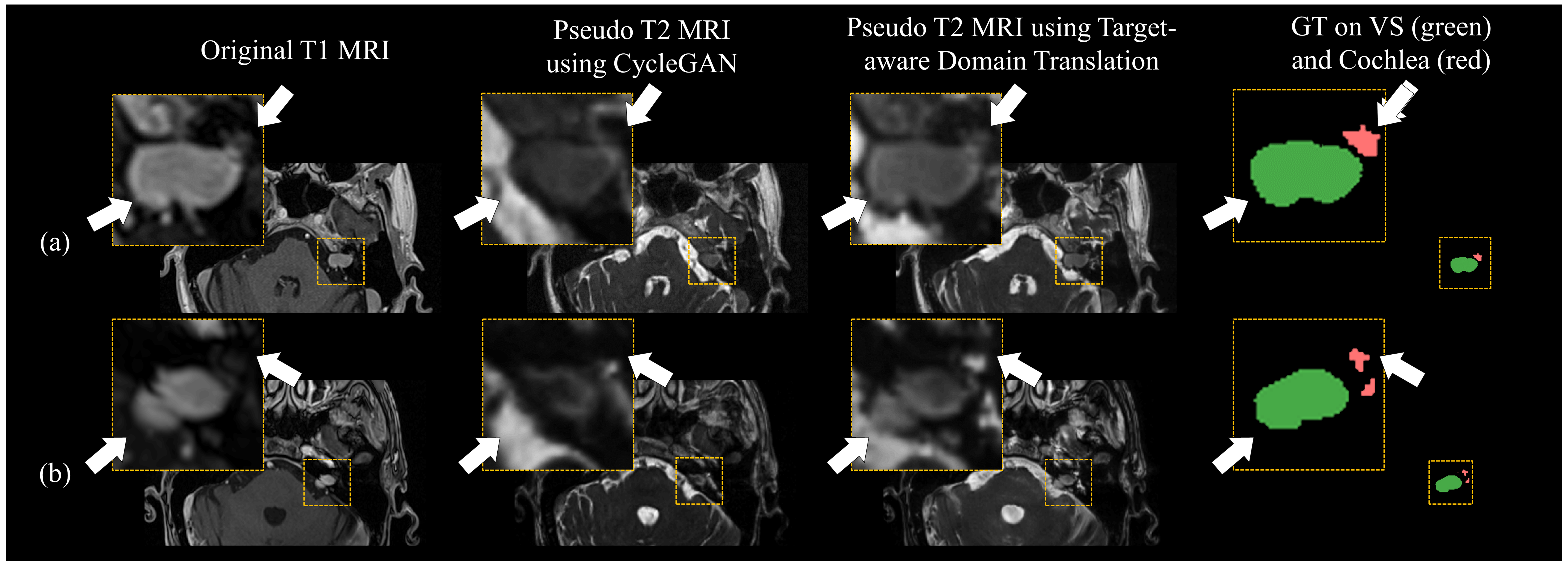}
\caption{ Qualitative comparison between the pseudo T2 MRI converted using target-aware domain translation and normal CycleGAN. From left, real T1 MRI, pseudo T2 MRI converted using CycleGAN, pseudo T2 MRI converted using target-aware domain translation network, and ground truth annotation on original T1 MRI are presented. As indicated by the magnified images and arrows, target-aware domain translation network preserves the shape of VS and cochlea, whereas CycleGAN fails to preserve the detailed structures. Best viewed in color and on high-resolution display.}
\label{fig:segmentor}
\hfill
\end{figure*}

There were many attempts that conducted end-to-end training of image translation and segmentation for domain adaptation. However, our approach differs in that, while the goal of the previous end-to-end image translation and segmentation approach is on unsupervised segmentation, our target-aware domain translation network aims to maintain the detailed shape of target structure-of-interest (i.e., VS and cochlea) through additional segmentation supervision and ultimately provide an accurate image translation. This proves to be effective in improving the performance of the segmentation network to be trained in the second phase (see \cref{fig:segmentor}, \Cref{tab:only}).

The proposed target-aware domain translation network in our COSMOS aims to train a network of an encoder $G_E^d$, a translation decoder $G_D^d$, a segmentation decoder $G_S^d$, and two discriminators $D^S$, $D^T$ that minimizes:
\begin{eqnarray}
\begin{aligned}
 & \mathcal{L}_{cosmos} ( G^{S \rightarrow T}, G^{T \rightarrow S} ,D^S, D^T) = \\
  &\lambda_{1} \mathcal{L}_{cycle}  (C^{S \rightarrow T}, C^{T \rightarrow S}) \\
+ &\lambda_{2} \mathcal{L}_{adv} (C^{S \rightarrow T}, C^{T \rightarrow S}, D^S, D^T) \\
+ &\lambda_{3} \mathcal{L}_{identity}(C^{S \rightarrow T}, C^{T \rightarrow S})  \\
+ &\lambda_{4} \mathcal{L}_{seg} (C^{S \rightarrow T}, Seg^{S \rightarrow T} , Seg^{T \rightarrow S}),  
\label{eq:1}
\end{aligned}
\end{eqnarray}

\noindent where $G^d =(G_E^d, G_D^d, G_S^d) $, domain converter $C^d = G_D^d \circ G_E^d $ , segmentor $ Seg^d = G_S^d \circ G_E^d $ for abbreviation, and domain  $ d \in \{ S \rightarrow T,T \rightarrow S \} $ for either source-to-target or target-to-source path. \\

\noindent \textbf{Cycle-consistency loss. }With the aid of CycleGAN structure, we impose L1 loss on cycle-consistency between scans of original source domain and those translated back from target to source domain, which is 
\begin{eqnarray}
\begin{aligned}
\mathcal{L}_{cycle} = & \left \| x^s  - C^{T \rightarrow S}(C^{S \rightarrow T}(x^s) \right \|_1 \\
                   +  & \left \| x^t  - C^{S \rightarrow T}(C^{T \rightarrow S}(x^t) \right \|_1
\label{eq:2}
\end{aligned}
\end{eqnarray}

\noindent \textbf{Adversarial loss. }Domain discriminator adversarial loss is 
\begin{eqnarray}
\begin{aligned}
\mathcal{L}_{adv} = \ &log(D^T(x^t)) + log(1-D^T(C^{S \rightarrow T}(x^s)) \\
                   + \ &log(D^S(x^S)) + log(1-D^S(C^{T \rightarrow S}(x^t))
\label{eq:3}
\end{aligned}
\end{eqnarray}

\noindent \textbf{Identity loss. }Identity loss is used in many CycleGAN based frameworks to preserve color composition between input and output by encouraging the translation network to be identity mapping when source and target samples are provided to target-to-source and source-to-target generators, respectively.
\begin{eqnarray}
\begin{aligned}
\mathcal{L}_{cycle} = & \left \| x^t  - C^{S \rightarrow T}(x^t) \right \|_1 \\
      +  & \left \| x^s  - C^{T \rightarrow S}(x^s) \right \|_1
\end{aligned}
\label{eq:4}
\end{eqnarray}

\begin{algorithm}
\caption{training process of the proposed method}\label{alg:cap}
\begin{algorithmic}[1]
\State Prepare ceT1 scans and its annotations $ (x^s,y^s) $, hrT2 scans $ x^t $. Initialize domain conversion network $ C^{S \rightarrow T} $, and teacher segmentation network $ f_{teacher}$

\State Train network $ C^{S \rightarrow T} $ with $ x^s $, $ y^s $ and $ x^t $
\State Convert $ x^s $ to pseudo hrT2 $\tilde{x}^t$ using $ C^{S \rightarrow T} $
\State Train teacher network $ f_{teacher} $ with labeled pseudo hrT2 scans $ (\tilde{x}^t, y^s) $.
\State Initialize first pseudo-label $ \tilde{y}^{(t, k=0)} = f_{teacher}({x}^t) $
\State Initialize student segmentation network $ f_{student} $, combined dataset of $ x^c = \{ \tilde{x}^t ,x^t\}$, $y^c = \{ y^s , \tilde{y}^{(t, k)} \} $ and iteration K

\For{$ k \gets 1 $  to $ K $}
    \State Train student $ f_{student} $ with $ \{ x^c , y^c \} $ 
    \State $ f_{teacher}  \gets  f_{student} $
    \State $ \tilde{y}^{(t, k-1)} = $ $ f_{teacher}(x^t) $
    \State $ y^c \gets \{ y^s, \tilde{y}^{(t, k-1)} \} $
    \EndFor
    
\end{algorithmic}
\label{alg:only}
\end{algorithm}

\noindent \textbf{Segmentor loss. } In addition to domain translation, we include segmentation training on both source-to-target and target-to-source direction. For source-to-target path, a shared encoder connected with both translation decoder and segmentation decoder learns features that satisfy both domain translation and segmentation, forcing translation network $ C^{S \rightarrow T} $ to be more focused on the target structures-of-interest (i.e., VS and cochlea). On the opposite path, this segmentor-included domain translation network performs segmentation on the synthetic target images $ \tilde{x}^t =C^{S \rightarrow T} (x^s) $, also regularizing the source-to-target domain translator to preserve the shape of structures-of-interest.
\begin{eqnarray}
\begin{aligned}
\mathcal{L}_{seg} = \ & (1-DSC(y^s, \ Seg^{S \rightarrow T}(x^s))) \ +\\
                    &  (1-DSC(y^s, \ Seg^{T \rightarrow S}( C^{S \rightarrow T} (x^s))))
\label{eq:5}
\end{aligned}
\end{eqnarray}

 



\subsection{Self-configuring network for segmentation from synthetic dataset}
\label{sec:subsec3_2}
As mentioned in section 2.2, configuring the model appropriately for the data characteristic heavily affects the model performance. Since our approach incorporate both labeled pseudo data and pseudo-labeled real data into training, it will be very difficult and laborious to empirically find the optimized model configuration (i.e., preprocessing, hyperparameters, model architecture, etc.) that fits our dataset. Therefore, we actively utilized the self-configuring pipeline of nnU-Net to find high quality configuration on our dataset. Details of the configuration are listed in \cref{sec.4.2}.

\subsection{Self-training}
\label{sec:3.3}
Even if the source-to-target transformed images are well generated, there may still exist some distribution gap with the real target domain data, and it is difficult to completely replace it. With nnU-Net as the backbone segmentation network, we proceeded with self-training scheme to incorporate non-annotated real target T2 scans into segmentation training. Our self-training scheme differs from the previous works on self-training in that the labeled data we use on self-training are in fact labeled pseudo-data (i.e., converted T2 scans paired with annotations provided with T1 scans) instead of real data.

Once nnU-Net is trained on labeled pseudo T2 scans, it is possible to get segmentation output on VS and cochlea by running real target T2 scans on the trained model. These segmentation results can be considered as pseudo-labels and used in the next step to include real target T2 scans (termed pseudo-labeled real target data hereafter) in segmentation training. Detailed description of self-training procedure is provided below.\\
\setlength \belowcaptionskip{-1ex}
\begin{table*}[t]
{\small
    \centering
    \begin{adjustbox}{width = \linewidth, center}
    \begin{tabular*}{\linewidth}{@{\extracolsep{\fill}}llllllll@{}}
    \toprule
                        &     & \multicolumn{3}{c}{\textbf{Dice coefficient} $\uparrow$ } & \multicolumn{3}{c}{\textbf{ASSD (mm)} $\downarrow$ } \\ \cmidrule(l){3-5} \cmidrule(l){6-8} 
       &  Methods / Team &
      \multicolumn{1}{c}{Mean} &
      \multicolumn{1}{c}{VS} &
      \multicolumn{1}{c}{Cochlea} &
      \multicolumn{1}{c}{Mean} &
      \multicolumn{1}{c}{VS} & \multicolumn{1}{c}{Cochlea} \\ \midrule

     & Team A \cite{MIP}         & 0.826$\pm$0.042    & 0.830$\pm$0.077 & 0.822$\pm$0.031 & 0.376$\pm$0.272  & 0.569$\pm$0.268    & 0.183$\pm$0.048     \\ 
     & Team B \cite{choi2021using}         & 0.830$\pm$0.047    & 0.838$\pm$0.083 & {\color[HTML]{00009B}\textbf{0.823$\pm$0.024}} & 0.733$\pm$1.036  & 1.294$\pm$1.237    & {\color[HTML]{00009B}\textbf{0.172$\pm$0.037}}   \\
     & Team C \cite{PAST}         & {\color[HTML]{00009B} \textbf{0.840$\pm$0.033}}    & {\color[HTML]{00009B}\textbf{0.870$\pm$0.066}} & 0.809$\pm$0.034  &{\color[HTML]{00009B}\textbf{0.298$\pm$0.211}}  & {\color[HTML]{FE0000}\textbf{0.418$\pm$0.242}}  & 0.177$\pm$0.045   \\ 
    \multirow{-4}{*}{\thead[l]{Comparative\\Results}}   
     & \textbf{COSMOS (ours) } &{\color[HTML]{FE0000} \textbf{0.857$\pm$0.034}} & {\color[HTML]{FE0000}\textbf{0.871$\pm$0.063}} & {\color[HTML]{FE0000}\textbf{0.842$\pm$0.020}}  & {\color[HTML]{FE0000}\textbf{0.294$\pm$0.238}}  &  {\color[HTML]{00009B}\textbf{0.437$\pm$0.270}}  &      {\color[HTML]{FE0000}\textbf{0.152$\pm$0.030}} \\      \midrule
      
    & w/o DA (source-only) & 0.209$\pm$0.197    & 0.181$\pm$0.330 & 0.237$\pm$0.203   & 23.44$\pm$22.02      & 33.91$\pm$23.58  & 9.920$\pm$8.650     \\
    & DA w/o Seg., w/o ST  & 0.721$\pm$0.213    & 0.715$\pm$0.282 & 0.727$\pm$0.177   & 3.911$\pm$9.823      & 5.905$\pm$12.62   & 1.835$\pm$5.134     \\
    & DA w/  Seg., w/o ST  & 0.800$\pm$0.127    & 0.770$\pm$0.238 & 0.830$\pm$0.037   & 0.468$\pm$0.793      & 0.767$\pm$1.056    & 0.177$\pm$0.068     \\
    & COSMOS (ST iter1)    & 0.842$\pm$0.054    & 0.851$\pm$0.099 & 0.834$\pm$0.023   & 0.357$\pm$0.371      & 0.490$\pm$0.394    & 0.164$\pm$0.038     \\
    & COSMOS (ST iter2)    & 0.853$\pm$0.034    & 0.864$\pm$0.065 & 0.842$\pm$0.024   & 0.311$\pm$0.261      & 0.469$\pm$0.294    & 0.154$\pm$0.031     \\
    \multirow{-6}{*}{\thead[l]{Ablation\\Results}} 
    & \textbf{COSMOS (ST iter3) }  &  \textbf{0.857$\pm$0.034} &       \textbf{0.871$\pm$0.063} &       \textbf{0.842$\pm$0.020}  & \textbf{0.294$\pm$0.238}    &        \textbf{0.437$\pm$0.270}       &       \textbf{0.152$\pm$0.030}
    \\ \bottomrule
    \end{tabular*}
    \end{adjustbox}
    \caption{Quantitative evaluation of the proposed method through comparative studies and ablation studies. COSMOS achieved the highest mean Dice value compared to other methods. 1st and 2nd placed results are colored in red and blue, respectively. In addition, the effectiveness of each module in the proposed method is demonstrated through the gradually improving performance in ablation studies.}
    \label{tab:only}
    }
\end{table*}

\setlength \belowcaptionskip{-3ex}

\noindent \textbf{Training segmentation with labeled pseudo T2 scans. }With the pseudo T2 scans $\tilde{x}^t$ converted from T1 scans and the annotations $y^s$ on the original T1 scans (i.e., labeled pseudo T2 dataset), we first train a teacher segmentation network $f_{teacher}$ that minimizes

\begin{equation}
\mathcal{L} = \frac{1}{n}\sum_{i}^{n}l_{seg}(y^s_i, f_{teacher}(\tilde{x}^t_i)).
\label{eq:6}
\end{equation}


\noindent \textbf{Inferring pseudo-labels by running non-annotated real target T2 scans to the trained model. }Even if our target-aware contrast conversion network in the earlier phase works well, there still exists domain gap between pseudo T2 and real T2 MR images. Therefore, performance drop may occur when real T2 data is passed to a model trained only with labeled pseudo T2 data. To alleviate this, we come up with the idea of combining two distributions together in the training set for segmentation. We inferred pseudo-labels $\tilde{y}^{(t,0)}$ of the non-annotated real T2 scans ${x}^t$ by running real T2 scans to the trained segmentation model $f_{teacher}$.

\begin{equation}
\tilde{y}^{(t,0)}_i = f_{teacher}( \{ x_i^t \}_{i=1}^{N_t}).
\label{eq:7}
\end{equation} 

\noindent \textbf{Retraining segmentation with combined data. }Pseudo T2 scans have different distribution with the real T2 scans, but they are paired with perfect annotations. On the other hand, real T2 scans are paired with pseudo-labels, which are yet incomplete. We incorporate both image-label pairs to self-training, to maximize the generalizability and minimize the performance degradation caused by difference in distributions. With the combined data of labeled pseudo T2 scans $ (\tilde{x}^t_i,y^s_i) $ and pseudo-labeled real T2 scans $ (x^t_i, \tilde{y}^{(t,0)}_i) $, we train a student segmentation network $f_{student}$ to minimize 

\begin{eqnarray}
\begin{aligned}
\mathcal{L}= \ &\frac{1}{n}\sum_{i}^{n}l_{seg}(y^s_i, f_{student}(\tilde{x}^t_i)) \\
    + &\frac{1}{m}\sum_{i}^{m}l_{seg}(\tilde{y}^{(t,0)}_i, f_{student}(x^t_i)))
\label{eq:8}
\end{aligned}
\end{eqnarray}\\
Despite not being ground truth labels, it has been reported in the previous literature that the self-training scheme and the pseudo-labels increases the performance and generalizability of the model on unseen data by utilizing unlabeled data into training \cite{xie2020self}.\\

\noindent \textbf{Iterating self-training procedure. }With self-training, we can include non-annotated real target data into training. By iterating self-training procedure (i.e., inferring new pseudo-labels by using the newly trained model and training another segmentation with the updated combined data), we can furthermore improve the quality of the pseudo-labels and expect improvement in the segmentation performance as the number of iteration $ k $ increases. We iterated self-training for a total of 3 iterations. The whole process of our proposed method is described in \Cref{alg:only} and \cref{fig2}.


%


\section{Experiments}

\subsection{Dataset}
All MRI scan data covered in this paper were acquired from a 32-channel Siemens Avanto 1.5T scanner using a Siemens single-channel head coil. Any necessary IRB approvals have been obtained \cite{shapey2021segmentation, shapey2021segmentation_medRxiv}. Contrast-enhanced T1-weighted imaging was performed with MPRAGE sequence of 0.4$\times$0.4mm$^2$ in-plane resolution, 512$\times$512 in-plane matrix, and 1.0 to 1.5mm slice thickness (TR=1900ms, TE=2.97ms, TI=1100ms). High-resolution T2-weighted scans were obtained with 3D CISS or FIESTA sequences of 0.5$\times$0.5mm$^2$ in-plane resolution, 384$\times$384 or 448$\times$448 in-plane matrix, 1.0-1.5 mm slice thickness (TR=9.4ms, TE=4.23ms). By deleting all health information identifiers and using a de-facing algorithm, data were completely de-identified. A total of 210 subject data that consists of 105 annotated ceT1 scans and 105 non-annotated hrT2 scans were used for training, and 32 non-annotated hrT2 scans were used for validation.

\begin{figure*}[h!]
\centering
\includegraphics[width=1\linewidth]{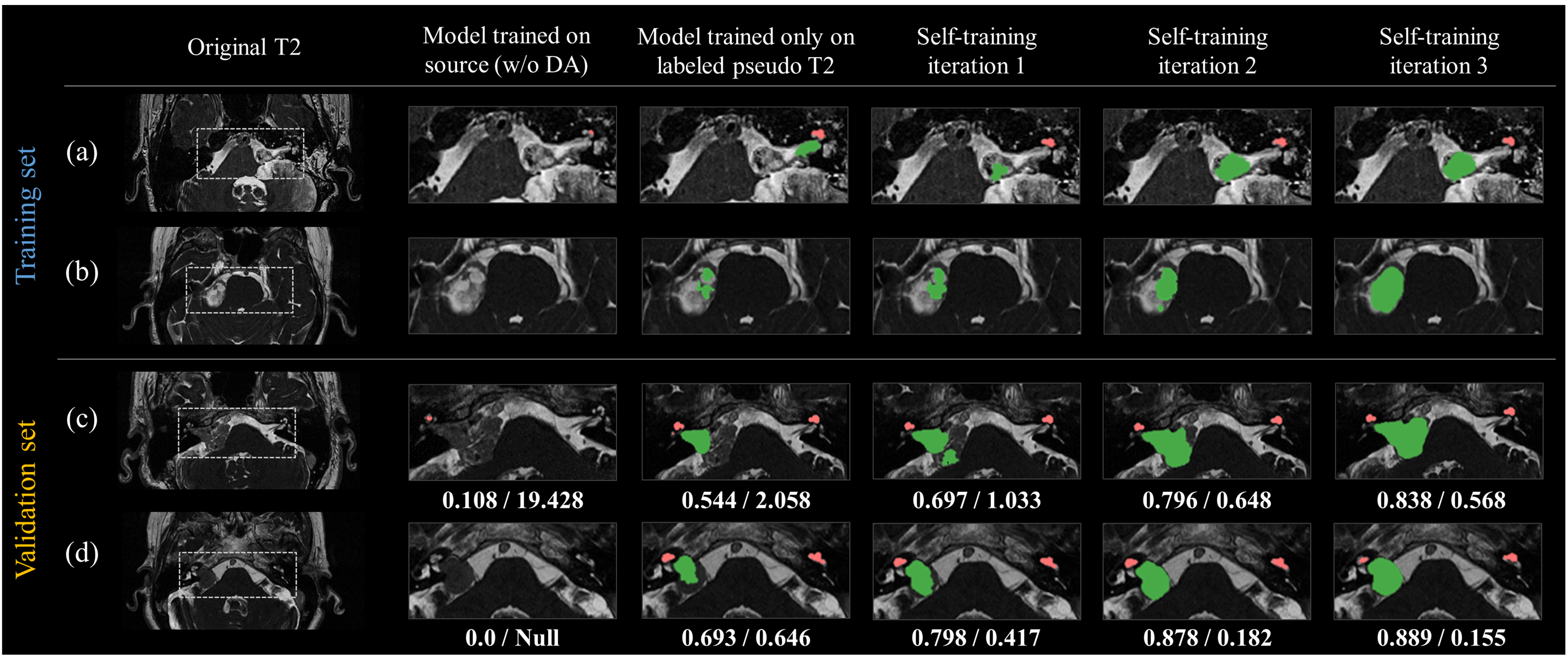}
\caption{Qualitative comparison of segmentation result as the proposed method is developed. Segmentation results on target training set (i.e., pseudo-labels) are presented on row (a, b), and results on validation set are presented on row (c, d). Evaluation metrics (i.e., mean Dice / ASSD) on each subject (c, d) in validation set are provided under figures. Please note that ground truth annotations on target T2 MRI were not provided, and quantitative metrics on validation set were acquired through official leaderboard. Best viewed in color and on high-resolution display.}
\label{fig:comb}
\hfill
\end{figure*}

\subsection{Implementation details}
\label{sec.4.2}
\noindent \textbf{Target-aware domain translation. }As mentioned in \cref{subsec:3_1}, our target-aware domain translation network follows the 2D CycleGAN structure, as two 3-level U-net based generators try to convert domains and two PatchGAN discriminators try to distinguish real scans from generated scans in each domain. Detailed structure is provided in the supplementary material. Based on this structure, we have attached segmentation networks that share the encoder of the generators and predict the VS and cochlea masks from input images. The segmentors are only trained when $ x^s $ and $\tilde{x}^{t}$ are fed to the generators since they are paired with the provided annotations $y^s$. Additionally, as there are few slices that contain annotations, segmentor loss was updated only when there exist foreground masks (i.e., VS and cochlea) in $ y^s $. Segmentation loss on $\tilde{x}^{t}$ was backpropagated after 5 epoch for stable training. We used min-max normalization for input images, and the model was trained for 50 epochs using Adam optimizer with an initial learning rate of 1e-4 and multiplicative decay. Weighting factors $\lambda_{1},\lambda_{2},\lambda_{3} $ and $\lambda_{4}$ in \eqref{eq:1} are set as 0.01, 0.1, 0.5 and 1 respectively.\\

\noindent \textbf{nnU-Net. } For every each step of self-training, 3D nnU-Net is initialized and trained to predict the next pseudo-labels. It includes deep supervision scheme with a total of 6 downsampling operations and initial number of 32 kernels. Each convolutional block consists of convolutional layers followed by instance normalization and leaky-ReLU activation. Training was done in 5-fold cross-validation. All data were preprocessed as follows: \\

\noindent \textbf{Cropping. } Cropped input patch size of our 3D nnU-Net was (40,224,224) for every batch with median patient size of (80,512,466), each representing z, y, x direction.  \\
\noindent \textbf{Resampling. } In-plane with third-order spline, out-of-plane with nearest neighbor, and target spacing for each subject was changed to (1.5, 0.410, 0.410).\\
\noindent \textbf{Intensity normalization. } 3D z-score normalization were applied for each subject. \\

For training, Dice and Cross Entropy loss were used with stochastic gradient descent optimizer of an initial learning rate 1e-2, batch size 2 for epoch 200. Data augmentation was applied on the fly. Detailed structure of the configured nnU-Net and data augmentations are provided in the supplementary material.

At inference step, test-time augmentation of flipping the input scan along all axis ($\times$8), and averaging the 8 predictions for each scan was applied. All-but-largest-suppression in 3D was applied to VS for post-processing. Predictions from each fold were ensembled to make final prediction. \\

\noindent \textbf{Self-training. }For effective self-training, we excluded pseudo-labels in which VS or cochlea were completely missing since those were considered as the ones that the model has low confidence on.


\subsection{Results}
\noindent \textbf{Target-aware translation. }Sample figures on the results of target-aware domain translation is shown in \cref{fig:segmentor}. For ablation study, we conducted domain translation without segmentors (which becomes pure CycleGAN) to prove the effectiveness of our segmentor-included domain translation. As observed in \cref{fig:segmentor}, pseudo T2 MRIs converted by our target-aware domain translation show better-preserved VS and cochlea compared to those converted by pure CycleGAN. Since the generated pseudo T2 MRIs are paired with labels on source T1 MRI, shape-preserved pseudo T2 MRI will lead to a segmentation model with improved performance. We observed that the segmentation model trained with shape-preserved pseudo T2 MRI shows better performance compared to that trained with shape-missing pseudo T2 MRI by a significant margin, as shown in \Cref{tab:only} (\textquotesingle DA w/o Seg.\textquotesingle \ vs. \textquotesingle DA w/ Seg.\textquotesingle).


\noindent \textbf{Results of dataset combination. } To see if incorporating both labeled pseudo T2 scans $ (\tilde{x}^t , y^s) $ and pseudo-labeled real T2 scans $ ( x^t, \tilde{y}^t) $ leads to an increase in segmentation performance and generalizability, we compared the segmentation performance of a model trained on combined dataset with that of a model trained only on labeled pseudo T2 scans. Under the identical experiment setting, including pseudo-labeled real T2 scans (i.e., applying self-training) showed increased performance (\Cref{tab:only}, \cref{fig:comb}). It is shown that Dice/ASSD metrics significantly improve when both domain data are incorporated. \\

\begin{figure}[h]
\centering
\begin{subfigure}[b]{0.5\linewidth}
\includegraphics[width=\textwidth]{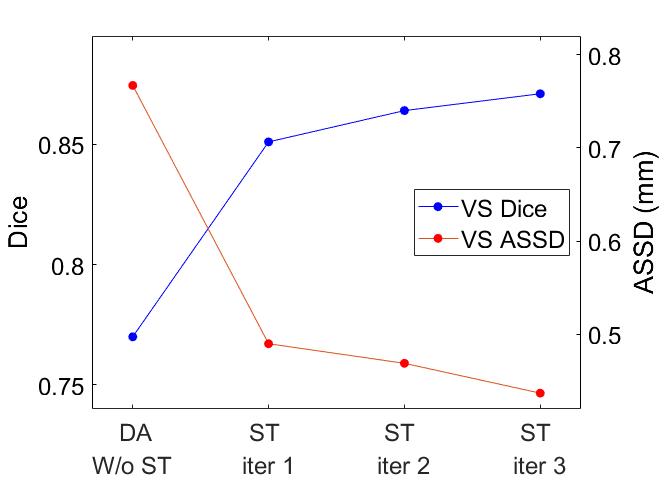}
\caption{VS Dice and ASSD}
\end{subfigure}\hfill
\begin{subfigure}[b]{0.5\linewidth}
\includegraphics[width=\textwidth]{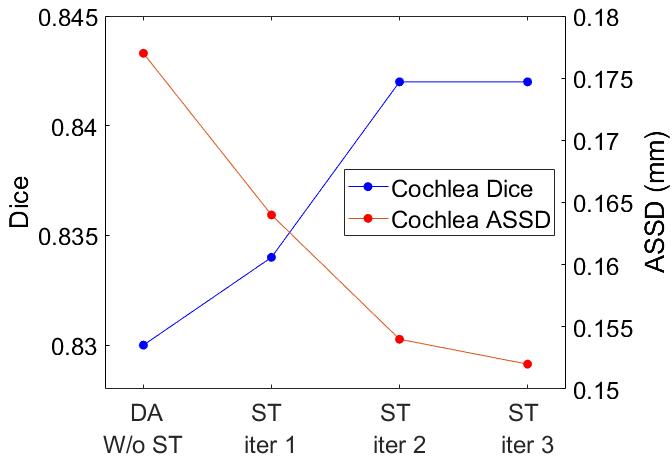}
\caption{Cochlea Dice and ASSD}
\end{subfigure}
\setlength \belowcaptionskip{-2ex}
\caption{Plot showing the gradual improvement for both VS and cochlea in Dice coefficient and ASSD.}
\label{fig:chart}
\hfill
\end{figure}

\noindent \textbf{Results of iterative pseudo-label updating. } We iterated self-training scheme to refine the pseudo-labels, thereby improving the segmentation model as the number of iteration increases. By setting the maximum number of iteration as three, we compared both quantitative and qualitative segmentation performance for segmentation models of each iteration. In \Cref{tab:only} and \cref{fig:chart}, it is shown that our iterative self-training scheme has gradually improved segmentation results on both VS and cochlea. In \cref{fig:comb}, we can see gradual improvement of segmentation on both training (i.e., pseudo-labels) and validation set, meaning our self-training scheme improves the imperfect pseudo-labels in training set as iteration increases. From \cref{fig:comb}.(a, b), it is shown that the pseudo-labels are correctly localized and filled as iterative self-training is performed. This also leads to incrementally increasing segmentation performance on validation set (\cref{fig:comb}.(c, d)). \\

\begin{figure}[t]
\centering
\includegraphics[width=1\linewidth]{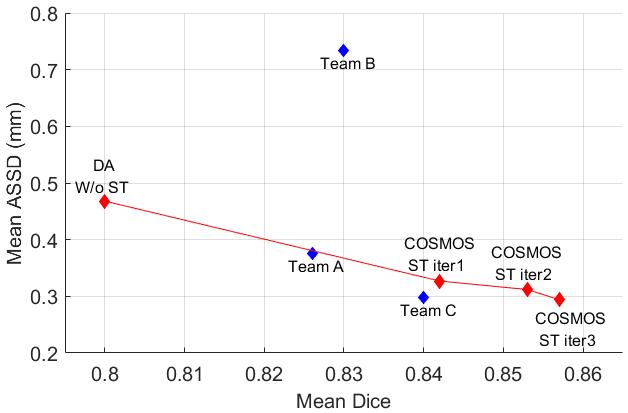}
\caption{Plot of final segmentation results of ours and other comparison methods. Our results with each iterations are colored in red, and we can see self-training incrementally improves the results.  COSMOS with self-training iteration 3 is placed at the most lower-right position, with highest mean Dice coefficient and lowest mean ASSD. }
\label{fig:pointplot}
\hfill
\end{figure}

\noindent \textbf{Comparison with other methods. } COSMOS won the 1\textsuperscript{st} place in the first Cross-Modality Domain Adaptation (crossMoDA) challenge \cite{dorent2022crossmoda} which aims to perform segmentation of VS and cochlea on hrT2 MR scans via UDA. Results of ours and three top-ranked methods \cite{PAST, choi2021using, MIP} in the challenge are provided in \cref{fig:pointplot} and \Cref{tab:only}. All teams follow the two-step approach of domain translation followed by segmentation. Team A \cite{MIP} and B \cite{choi2021using} added contrastive learning to the domain translation process, but they didn't apply self-training in segmentation phase. Instead, Team A utilized Mean Teacher \cite{tarvainen2017mean} as semi-supervised approach for segmentation. Team C \cite{PAST} uses separate models for each image size (384$\times$384 or 448$\times$448) and target structure (i.e., VS or cochlea), which is attributed as the reason for the slightly lower ASSD of VS segmentation performance than ours. However, they show relatively poor cochlea segmentation performance due to the lack of shape-preserving regularization in the domain translation process, without which the cochlea can easily disappear in the transformed images. \cref{fig:pointplot} shows that our result is placed at the most lower-right position of the chart with state-of-the-art performance in Dice coefficient and ASSD, improving in both metrics as the number of iteration increases.

\section{Discussion}
\label{discussions}
In this work, we verified COSMOS in unsupervised domain adaptation from source contrast-enhanced T1 domain to target high-resolution T2 domain for 3D semantic segmentation of VS and cochlea. Through this, a much lower-cost and safer alternative can be provided to the patient in the clinical process of follow-up and treatment planning. A current limitation of COSMOS for unsupervised medical image segmentation is that it has not yet been validated in various medical datasets. We will develop COSMOS as a generalizable methodology in unsupervised medical segmentation tasks by extending COSMOS to various medical datasets through future work.

Self-training has recently achieved great success in improving the performance of classifiers through a semi-supervised approach in tasks such as image classification and speech recognition \cite{yalniz2019billion, xie2020self, kahn2020self}. In self-training for classification, it is known that the performance is improved if the images that the model has low confidence on are filtered in the process of forming pseudo-labeled data. Applying self-training to unsupervised segmentation, we referred to this and filtered the unlabeled images that the model has missed VS or cochlea on the segmentation output. In addition to this image-level filtering, we aim to develop a more accurate and generalizable methodology by pixel-level filtering in which a pseudo-supervision is given only to pixels with high confidence and expand high-confident pseudo-label area through iteration.

\setlength \belowcaptionskip{-4ex}

\section{Conclusion}
\label{conclusion}
In this study, we proposed cross-modality domain adaptation for unsupervised 3D medical image segmentation based on target-aware domain translation and self-training named COSMOS, and validated it on unsupervised segmentation of VS and cochlea on high-resolution T2 scans. Various ablation studies demonstrate the effectiveness of each module in COSMOS, and quantitative comparison with other methods shows that COSMOS achieves state-of-the-art result in unsupervised VS and cochlea segmentation. COSMOS can be extended to various unsupervised 3D medical image segmentation applications to improve the applicability of deep learning approaches in clinical settings. \\ \\

\noindent \textbf{\large{Acknowledgements}} \\
\indent This research was supported by Basic Science Research Program through the National
Research Foundation of Korea (NRF) funded by the Ministry of Science and ICT (2019R1A2B5B01070488, 2021R1A4A1031437, 2021R1C1C2008773), Brain Research Program through the NRF funded by the Ministry of Science, ICT \& Future Planning (2018M3C7A1024734), Y-BASE R\&E Institute a Brain Korea 21, Yonsei University, and the Artificial Intelligence Graduate School Program, No. 2020-0-01361, Yonsei University.

{\small
\bibliographystyle{ieee_fullname}
\bibliography{egbib}
}

\end{document}